\documentclass[12pt,epsfig]{article}
\usepackage{epsfig}
\begin{document}

\begin{center} 
{\Large {\bf Gauge bosons masses in the context of the Supersymmetric 
$SU(3)_{C}\otimes SU(3)_{L}\otimes U(1)_{N}$ Model}}
\end{center}

\begin{center}
M. C. Rodriguez  \\
{\it Grupo de F{\'{\i}}sica Te\'{o}rica e Matem\'{a}tica F\'{\i}sica \\
Departamento de F\'{\i}sica  \\
Universidade Federal Rural do Rio de Janeiro - UFRRJ \\
BR 465 Km 7, 23890-000, Serop\'{e}dica - RJ \\
Brasil}
\end{center}

\date{\today}

\begin{abstract}
We analyze the boson masses and their mixing in the Minimal Supersymmetric 
$SU(3)_{C}\otimes SU(3)_{L}\otimes U(1)_{N}$
Model, and we will show all the numerical results are in 
agreement with actual current experimental limits.
\end{abstract}

PACS number(s): 12.60. Jv



\section{Introduction}

Recently the CDF Collaboration at Fermilab presented its new High-precision measurement, with an accuracy of $\sim 10^{-4}$, of the $W$-boson mass \cite{cdf}
\begin{equation}
\left( M_{W} \right)_{CDF}= \left(
80.4335 \pm 0.0094 \right) {\mbox GeV},
\label{cdfresult}
\end{equation}
this measurement represent an excess bigger than 
$6$ $\sigma$ in relation to the more precise value ({\it conservative} scenario) of 
the Standard Model (SM) \cite{sg,Kronfeld:2010bx} which is given by \cite{average}
\begin{equation}
\left( M_{W} \right)_{SM}= \left(
80.3505 \pm 0.0077 \right) {\mbox GeV}.
\label{smvalue}
\end{equation}
Clearly if this result, presented by the CDF, is confirmed 
by other experimental collaborations, this mean a new 
indication of physics beyond the SM.

From the theoretical point of view, the SM cannot be a fundamental theory since it has so many 
questions like that of the number of families do not have an answer in its context. One of 
these  possibilities to solve this problem is that, at energies of a few TeVs, the gauge symmetry may be 
\begin{equation} 
SU(3)_{C}\otimes SU(3)_{L} \otimes U(1)_{N},
\end{equation} 
(or it is more known as M331 for shortness) instead of that of 
the SM \cite{pp,pf}. They are interesting  possibilities for 
the physics at the TeV scale and we can also accomodate 
the CDF data \cite{VanLoi:2022eir}. When 
$\sigma_{1}$ get non zero value the neutrinos get the 
following Majorana mass term \cite{Pisano:1997hk}
\begin{equation}
\frac{G_{ab}}{2 \sqrt{2}}< \sigma^{0}_{1}>.
\end{equation}

The supersymmetric version of the 331 minimal model, or it 
is more known as MSUSY331 for shortness, was considered \cite{ema1,pal2,331susy1,mcr}. In 
this model beyond 
the anti-sextet $S$, we need to introduce a Sextet  
$S^{\prime}$ Higgs boson, to  cancel chiral anomalies 
generated by the superpartners of $S$ in similar way as 
we have to add a new doublet scalar at Minimal 
Supersymmetric Standard Model (MSSM). 

When all the neutral fields in sextet and anti-sextet get vev, 
the neutrinos obtain mass via the see-saw mechanism, due to 
the mix between higgsinos with the usual leptons, which is 
given by:
\begin{equation}
\mu_{0i}\hat{L}_{i}\hat{\eta}^{\prime}+
\lambda_{2ij} (\epsilon \hat{L}_{i}\hat{L}_{j}\hat{\eta})+
\lambda_{3ij} (\hat{L}_{i}\hat{S}\hat{L}_{j}) +
\lambda_{4i}(\epsilon\hat{L}_{i}\hat{\chi}\hat{\rho}), 
\end{equation}
in a similar way to what we have done \cite{Montero:2001ch}. It must be explore it in the future. 

The vaccum expectation 
value of this new scalar field, $S^{\prime}$, can expain the shift on the $W$ mass as we will shown below.
We can decompose our scalars fields in 
$SU(2)\otimes U(1)$ representations \cite{Liu:1993gy,Montero:1999mc,DeConto:2015eia}
\begin{eqnarray}
{\bf 6}_{0} \rightarrow 
{\bf 3}_{2}\oplus 
{\bf 2}_{-1} \oplus {\bf 1}_{-4}
\end{eqnarray}
where
\begin{eqnarray}
S&=& \left( 
\begin{array}{cc} 
T & \frac{\Phi_{S}}{\sqrt{2}} \\ 
\frac{\Phi^{T}_{S}}{\sqrt{2}} & H^{--}_{2}          
\end{array} \right)
\sim ({\bf 1},{\bf 6}^{*},0), 
\nonumber \\
T&=&\left( \begin{array}{cc} 
\sigma^{0}_{1} & \frac{h^{+}_{1}}{\sqrt{2}} \\ 
\frac{h^{+}_{1}}{\sqrt{2}} & H^{--}_{1} \\
\end{array} \right)
\sim ({\bf 1},{\bf 3},+2), \nonumber \\
\Phi_{S}&=& \left( 
\begin{array}{c} 
h^{+}_{2} \\ 
\sigma^{0}_{2}          
\end{array} \right)
\sim ({\bf 1},{\bf 2},-1), \quad 
H^{--}_{2}
\sim ({\bf 1},{\bf 1},-4).
\label{sig1:triplet} 
\end{eqnarray}
$\Phi_{S}$ is the Higgs doublet of SM while $T$ was 
considered at \cite{Gelmini:1980re}. One 
of main goal at Large Hadron Collider (LHC) is to find new physics beyond the 
SM. Then it is useful to 
predict the masses of new particles arising in some interesting 
models. We want to 
present a detailed analyses of the gauge bosons masses in 
similar way as we have done in the scalar sector 
\cite{Rodriguez:2021dnb,Rodriguez:2005jt}. We have also presented a previous mass spectrum of 
this model, without the Sextet \cite{mcr,Rodriguez:2010tn}, in this article we have also studied 
the processes
\begin{equation}
g,d \rightarrow U^{--}+J, \,\ g,u \rightarrow U^{--}+j_{\alpha}.
\end{equation}
These exotic quarks have charge $(5/3)e$ and $-(4/3)e$, respectivelly. These processes 
was suggested by Alexandre Belyaev and its signature is $llX$ and it can be detected at 
LHC if they really exist in nature.

We will present in next Section, Sec.(\ref{sec:analitico}), the Minimal Supersymmetric 331 
Model, (MSUSY331), and our numerical results are present at 
Sec.(\ref{wcdf}). Our conclusions are presented at the last Section 
of this article. We present in appendices the analytical analysis of the 
mixtures of the gauge bosons of this model and some numerical analyses about the mixing in 
tne neutral boson sector..

\section{Minimal Supersymmetric 331 Model (MSUSY331)} 
\label{sec:analitico}

The scalars in this model are given by \cite{331susy1,mcr}:
\begin{eqnarray}
\eta &=& 
\left( \begin{array}{c} 
\eta^{0} \\ 
\eta^{-}_{1} \\
\eta^{+}_{2}          
\end{array} \right) 
\sim ({\bf 1},{\bf 3},0),\quad
\rho = 
\left( \begin{array}{c} 
\rho^{+} \\ 
\rho^{0} \\
\rho^{++}          
\end{array} \right) 
\sim ({\bf 1},{\bf 3},+1), \nonumber \\ 
\chi &=& 
\left( \begin{array}{c} 
\chi^{-} \\ 
\chi^{--} \\
\chi^{0}          
\end{array} \right) 
\sim ({\bf 1},{\bf 3},-1), \,\  
\tilde{\chi}^{\prime} = 
\left( \begin{array}{c} 
\tilde{\chi}^{\prime+} \\ 
\tilde{\chi}^{\prime++} \\
\tilde{\chi}^{\prime0}          
\end{array} \right) 
\sim ({\bf1},{\bf3}^{*},+1), \nonumber \\
\tilde{\eta}^{\prime} &=& 
\left( 
\begin{array}{c} 
\tilde{\eta}^{\prime0} \\ 
\tilde{\eta}^{\prime+}_{1} \\
\tilde{\eta}^{\prime-}_{2}          
\end{array} \right) 
\sim ({\bf1},{\bf3}^{*},0),\quad
\tilde{\rho}^{\prime} = 
\left( \begin{array}{c} 
\tilde{\rho}^{\prime-} \\ 
\tilde{\rho}^{\prime0} \\
\tilde{\rho}^{\prime--}          
\end{array} 
\right) 
\sim ({\bf1},{\bf3}^{*},-1), \nonumber \\ 
S &=& \left( \begin{array}{ccc} 
\sigma^{0}_{1}& 
\frac{h^{+}_{2}}{ \sqrt{2}}& \frac{h^{-}_{1}}{ \sqrt{2}} \\ 
\frac{h^{+}_{2}}{ \sqrt{2}}& H^{++}_{1}& \frac{ \sigma^{0}_{2}}{ \sqrt{2}} \\
 \frac{h^{-}_{1}}{ \sqrt{2}}& 
\frac{\sigma^{0}_{2}}{ \sqrt{2}}&  H^{--}_{2}        
\end{array} \right) \sim ({\bf1},{\bf6}^{*},0), \nonumber \\
\tilde{S}^{\prime} &=& 
\left( \begin{array}{ccc} 
\tilde{\sigma}^{\prime0}_{1}& \frac{\tilde{h}^{\prime-}_{2}}{ \sqrt{2}}& 
\frac{\tilde{h}^{\prime+}_{1}}{ \sqrt{2}} \\ 
\frac{\tilde{h}^{\prime-}_{2}}{ \sqrt{2}}& \tilde{H}^{\prime--}_{1}& 
\frac{ \tilde{\sigma}^{ \prime 0}_{2}}{ \sqrt{2}} \\
 \frac{\tilde{h}^{\prime+}_{1}}{ \sqrt{2}}& 
\frac{ \tilde{\sigma}^{ \prime 0}_{2}}{ \sqrt{2}}&  
\tilde{H}^{\prime++}_{2}        
\end{array} \right) \sim ({\bf1},{\bf6},0).
\label{escoriginais} 
\end{eqnarray}

The charged gauge bosons of this model are:
\begin{eqnarray}
W^{ \pm}_{m}(x)&=&-\frac{1}{\sqrt{2}}(V^{1}_{m}(x) 
\mp \imath V^{2}_{m}(x)),
\nonumber \\
V^{ \pm}_{m}(x)&=&-\frac{1}{\sqrt{2}}(V^{4}_{m}(x) 
\pm \imath V^{5}_{m}(x)), 
\nonumber \\
U^{\pm \pm}_{m}(x) &=&- \frac{1}{\sqrt{2}}(V^{6}_{m}(x) 
\pm \imath V^{7}_{m}(x)).
\label{defcarbosons}
\end{eqnarray}
while the neutral gauge bosons are the photon
\begin{eqnarray}
A_{m}&=& \sin \theta_{W} \left( V^{3}_{m}- \sqrt{3}V^{8}_{m} \right) + 
\sqrt{1-4 \sin^{2} \theta_{W}}V_{m},
\end{eqnarray}
where $\theta_{W}$ is the Weinberg angel and we have also two massives neutral bosons defined as
\begin{eqnarray}
Z_{m}&=& \cos \theta_{W}V^{3}_{m}+ \sqrt{3}\tan \theta_{W}\sin \theta_{W}V^{8}_{m}- 
\tan \theta_{W}\sqrt{1-4 \sin^{2} \theta_{W}}V_{m}, \nonumber \\
Z^{\prime}_{m}&=& \frac{1}{\cos \theta_{W}}\left[
\sqrt{1-4 \sin^{2} \theta_{W}}V^{8}_{m}+ \sqrt{3}\sin \theta_{W}V_{m} \right].
\end{eqnarray}

In order to get the gauge bosons masses we have to calculate
\begin{eqnarray}
{\cal L}^{Escalar}_{Higgs}&=& ({\cal D}_{m} \eta)^{\dagger}({\cal D}^{m} \eta)+ 
({\cal D}_{m} \rho)^{\dagger}({\cal D}^{m} \rho)+
({\cal D}_{m} \chi)^{\dagger}({\cal D}^{m} \chi)+
(\overline{{\cal D}_{m}} \eta^{\prime})^{\dagger}(\overline{{\cal D}^{m}} \eta^{\prime}) \nonumber \\ &+&
(\overline{{\cal D}_{m}} \rho^{\prime})^{\dagger}(\overline{{\cal D}^{m}} \rho^{\prime})+
(\overline{{\cal D}_{m}} \chi^{\prime})^{\dagger}(\overline{{\cal D}^{m}} \chi^{\prime})+
Tr[({\cal D}_{m}S)^{\dagger}({\cal D}^{m}S)] \nonumber \\
&+&
Tr[(\overline{{\cal D}_{m}}S^{\prime})^{\dagger}(\overline{{\cal D}^{m}}S^{\prime})]. \nonumber \\
\label{lagHHVV}
\end{eqnarray}

The  mass of charged gauge boson are given by \cite{mcr,Rodriguez:2010tn} 
\begin{eqnarray}
M^{2}_{W}&=& \frac{g^{2}}{4} \left[
v^{2}_{ \eta}+v^{2}_{ \rho}+v^{2}_{ \sigma_{2}}+v^{2}_{ \eta^{\prime}}+v^{2}_{ \rho^{\prime}}+ 
2 \left( v^{2}_{\sigma_{1}}+v^{2}_{\sigma^{\prime}_{1}} \right) +
v^{2}_{\sigma^{\prime}_{2}} \right], \nonumber \\
M^{2}_{V}&=& \frac{g^{2}}{4}\left[
v^{2}_{ \eta}+v^{2}_{ \chi}+v^{2}_{ \sigma_{2}}+v^{2}_{ \eta^{\prime}}+v^{2}_{ \chi^{\prime}}+ 
2 \left( v^{2}_{\sigma_{1}}+v^{2}_{\sigma^{\prime}_{1}} \right) +
v^{2}_{\sigma^{\prime}_{2}} \right], \nonumber \\
\delta_{WV}&=&\frac{g^{2}}{\sqrt{2}}\left(
v_{\sigma_{1}}v_{\sigma_{2}}+ 
v_{\sigma^{\prime}_{1}}v_{\sigma^{\prime}_{2}} 
\right), \nonumber \\
M^{2}_{U}&=& \frac{g^{2}}{4} \left(
v^{2}_{\rho}+v^{2}_{\chi}+4v^{2}_{\sigma_{2}}+
v^{2}_{\rho^{\prime}}+v^{2}_{\chi^{\prime}}+
4v^{2}_{\sigma^{\prime}_{2}} 
\right). 
\label{bgmcc}
\end{eqnarray}
This new charged vector boson, $V^{\pm}$, is best known 
in the literature as being $W^{\prime}$ and in recent analyzes its masses have been considered in the following range \cite{osland} 
\begin{equation}
0.25 \leq M_{W^{\prime}} \leq 2 \,\ {\mbox TeV}.
\label{explimV}
\end{equation}
The charged gauge bosons can mixing. We present the analytical analysis of this mixture in 
Sec.(\ref{sec:mixingchar}). We also have, as in SM, the usual $Z$-boson and an extra neutral 
gauge boson known as $Z^{\prime}$-boson and in the approximation that 
$v_{ \chi} \simeq v^{2}_{ \chi^{\prime}}$ and they are bigger than the others vev of this 
model we get the following values \cite{mcr}
\begin{eqnarray}
M^{2}_{Z}&\approx& \frac{1}{2} 
\left( \frac{g^{2}+4g^{\prime 2}}{g^{2}+3g^{\prime 2}} \right)
\left( v^{2}_{ \eta}+v^{2}_{ \rho}+v^{2}_{ \sigma_{2}}+v^{2}_{ \eta^{\prime}}+v^{2}_{ \rho^{\prime}}+ v^{2}_{\sigma^{\prime}_{2}} \right), \nonumber \\
M^{2}_{Z^{\prime}}&\approx& \frac{2}{3} (g^{2}+3g^{\prime 2}) 
(v^{2}_{ \chi}+v^{2}_{ \chi^{\prime}}),
\label{zmassmcr}
\end{eqnarray}
using the first expression from Eq.(\ref{bgmcc}), toghether with $M^{2}_{Z}$, we can write
\begin{equation}
\frac{M^{2}_{Z}}{M^{2}_{W}}= \frac{1+4t^{2}}{1+3t^{2}},
\label{tinthetaw}
\end{equation}
where. we have defined the new parameter $t$ in the following way:
\begin{equation}
t \equiv \frac{g^{\prime}}{g},
\end{equation}
but from the SM we know the following result
\begin{equation}
\frac{M^{2}_{Z}}{M^{2}_{W}}= \frac{1}{1- \sin^{2} \theta_{W}},
\label{tinthetaw}
\end{equation}
where $\theta_{W}$ is the Weak mixing angle. When we impose Eq.(\ref{tinthetaw}) is the same 
results are presented in Eq.(\ref{tinthetaw}), we get the famous relation
\begin{equation}
t^{2}= \frac{\sin^{2} \theta_{W}}{1-4 \sin^{2} \theta_{W}},
\label{tinthetaw2}
\end{equation}
It imply
\begin{eqnarray}
\sin^{2}\theta_{W}< \frac{1}{4},
\end{eqnarray} 
with a Landau pole
\footnote{Exists an energy scale $\mu$ where the model loses its perturbative regime, see \cite{Dias:2004dc}.} in 
\begin{eqnarray}
\sin^{2}\theta_{W}(\mu)= \frac{1}{4},
\end{eqnarray}
this result is in agreement with the actual experimental bound, given by  
\begin{equation}
\sin^{2}\theta_{W}=1- \frac{M^{2}_{W}}{M^{2}_{Z}}=0.223562,
\label{expsintw}
\end{equation}
our neutral gauge bosons can mixing, the analytical results without this approximation 
is presented at Sec.(\ref{sec:mixingneu}). The limits on the $Z^{\prime}$ mass is given by \cite{osland} 
\begin{equation}
1.0 \leq M_{Z^{\prime}} \leq 4.5 \,\ {\mbox TeV}.
\label{explimZP}
\end{equation}

\section{Numerical Analyses for CDF Experimental Results.}
\label{wcdf}

The masses of the $W$-boson and $Z$-boson receive 
a new tree-level contribution given by:
\begin{eqnarray}
\delta M^{2}_{W}&=& \frac{g^{2}}{4}\left[ 2 \left(
v^{2}_{\sigma^{1}}+v^{2}_{\sigma^{\prime}_{1}} \right) +
v^{2}_{\sigma^{\prime}_{2}} \right], \,\
\delta M^{2}_{Z}= \frac{g^{2}}{4 \cos^{2}\theta_{W}} \left[ 4 \left(
v^{2}_{\sigma^{1}}+v^{2}_{\sigma^{\prime}_{1}} \right) +
v^{2}_{\sigma^{\prime}_{2}} \right], \nonumber \\
\label{resmsusy331}
\end{eqnarray}
therefore, as first prevision of this model is
\begin{equation}
\delta M^{2}_{W} \neq \delta M^{2}_{Z},
\end{equation}
our analytical results are in agreement with the 
results obtained for the M331 as presented by the references 
\cite{Pisano:1997hk,Montero:1999su,Sharma:1999ct}. In order, 
to explain the $W$-mass anomaly we have to impose:
\begin{equation}
\left| \sqrt{2 \left( v^{2}_{\sigma_{1}}+v^{2}_{\sigma^{\prime}_{1}}\right) +
v^{2}_{\sigma^{\prime}_{2}}} \right|= 
\frac{2}{g}
\sqrt{| \left( M_{W} \right)^{2}_{SM}-
\left( M_{W} \right)^{2}_{CDF}|} 
=11.19 {\mbox GeV}.
\label{wmassrest}
\end{equation}
This new contribution does not have 
any restrictions coming from the fermion mass explanation. 

Recently it was proposed this shift can be easily explained by a real triplet Higgs 
boson \cite{Blank:1997qa,Chen:2006pb} and the triplet vev, 
$v_{T}$, is around $10$ GeV. We can also use an extra complex triplet Higgs boson with $v_{T}\approx 3.2$ GeV \cite{evans} to expain this anomaly. There is also, an interesting proposal to 
explain this anomaly using the Minimal $R$-Symmetric extension of MSSM, known as 
MRSSM \cite{Kribs:2007ac,Diessner:2014ksa,Diessner:2019ebm} and in this case $|v_{T}|\leq 4$ GeV \cite{Diessner:2014ksa}. The mechanism is based on the 
fact that in this model a new scalar field is introduced in the Triplet representation and it 
is not necessary to generate mass for the fermions \cite{Diessner:2014ksa,Diessner:2019ebm}, 
the vev of this new scalar can be around $3$ GeV \cite{athron}. 

By another hand, for the $\rho$-parameter, in our model, it 
is hold the following expression:
\begin{eqnarray}
\rho &=& \left( 
\frac{M^{2}_{W}}{M^{2}_{Z}\cos^{2}\theta_{W}} \right)=
\frac{1+R}{1+R^{\prime}}, \nonumber \\
R&=& 2 \left( \frac{v^{2}_{\sigma_{1}}+v^{2}_{\sigma^{\prime}_{1}}}{v^{2}_{MP}} \right) +
\frac{v^{2}_{\sigma^{\prime}_{2}}}{v^{2}_{MP}}, \,\
R^{\prime}= 4 \left( \frac{v^{2}_{\sigma_{1}}+v^{2}_{\sigma^{\prime}_{1}}}{v^{2}_{MP}} \right) +
\frac{v^{2}_{\sigma^{\prime}_{2}}}{v^{2}_{MP}}, \nonumber \\
\rho &=&1+R-R^{\prime}=1-2 
\left( \frac{v^{2}_{\sigma_{1}}+v^{2}_{\sigma^{\prime}_{1}}}{v^{2}_{MP}} \right).
\end{eqnarray} 
remember the experimental measurement of this parameter is 
given by: $\rho =0.9998\pm 0.0008$, then we must satisfy the 
following inequality:
\begin{eqnarray}
\sqrt{ v^{2}_{\sigma_{1}}+v^{2}_{\sigma^{\prime}_{1}}} < 
\sqrt{\frac{0.0008}{2}}v_{MP}<5 {\mbox GeV}.
\label{rhorest}
\end{eqnarray}

So we can explain the new CDF data for the $W$-mass, see 
Eq.(\ref{wmassrest}), and the $\rho$-parameter, 
Eq.(\ref{rhorest}), if the new vev of this model are of the 
order of a few GeV. First result can be taken all the vev 
from the triplet equal to zero, it means
\begin{equation}
v^{2}_{\sigma^{\prime}_{1}}=v^{2}_{\sigma^{\prime}_{1}}=0 \,\ 
{\mbox GeV},
\end{equation}
it imply $\rho =1$ and also
\begin{eqnarray}
\delta M^{2}_{Z}&=&\frac{\delta M^{2}_{W}}{\cos^{2}\theta_{W}} \Rightarrow 
\delta M^{2}_{Z}=17.187592 \,\ {\mbox GeV}^{2}, \nonumber \\
v_{\sigma^{\prime}_{2}}&=&11.1942 \,\ {\mbox GeV}.
\label{mzhip1}
\end{eqnarray}

When we take the 
\begin{equation}
v^{2}_{\sigma^{\prime}_{1}}=v^{2}_{\sigma^{\prime}_{1}}=0 \,\ 
{\mbox GeV},
\end{equation}
our Eq.(\ref{rhorest}) give the upper limit $v^{2}_{\sigma_{1}}<5$ GeV 
the same result given in \cite{Montero:1999su}, but in order to also 
satisfy Eq.(\ref{wmassrest}), we must have
\begin{equation}
v_{\sigma_{1}}=7.91552 \,\ {\mbox GeV},
\end{equation}
therefore, this solution is not possible inside m331 model, but it 
is possible in MSUSY331.

\begin{figure}[ht]
\begin{center}
\vglue -0.009cm
\mbox{\epsfig{file=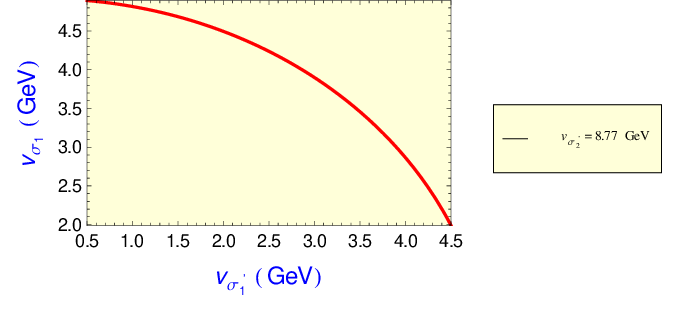,width=0.7\textwidth,angle=0}}       
\end{center}
\caption{Possible values for $v_{\sigma_{1}}$ as function 
of $v_{\sigma^{\prime}_{1}}$, in the minimal supersymmetric  
$SU(3)_{C}\otimes SU(3)_{L}\otimes U(1)_{N}$ Model,  see 
Eq.(\ref{sigma1VSsigma1p}).}
\label{fig6}
\end{figure}

We must have all new vev different of zero. We can use 
Eqs.(\ref{wmassrest},\ref{rhorest}) for fix $v_{\sigma^{\prime}_{2}}$:
\begin{eqnarray}
v_{\sigma^{\prime}_{2}}&=& 
\sqrt{\frac{4\delta M^{2}_{W}}{g}-erro*v^{2}_{MP}}=
8.7691841 \,\ {\mbox GeV}.
\label{hip3}
\end{eqnarray}
Using Eq.(\ref{rhorest}) we can define $v_{\sigma_{1}}$:
\begin{equation}
v_{\sigma_{1}}= \sqrt{\frac{2\delta M^{2}_{W}}{g}- 
\frac{v^{2}_{\sigma^{\prime}_{2}}}{2}-v^{2}_{\sigma^{\prime}_{1}}}
\label{sigma1VSsigma1p}
\end{equation}
our results is shown in Fig.(\ref{fig6}). Now we get the following 
numerical results:
\begin{eqnarray}
\rho &=&0.9992, \nonumber \\
\delta M^{2}_{Z}&=&28.8278 \,\ {\mbox GeV}^{2}, \,\ \Rightarrow 
\left( M_{Z} \right)_{CDF}=91.3181 \,\ {\mbox GeV} .
\label{mzhip3}
\end{eqnarray} 

We will consider, for the charged gauge bosons, the following two case:
\begin{itemize}
\item[a-)] Only $v_{\sigma^{\prime}_{2}} \neq 0$, in this case the gauge bosons $W$ and 
$V$ are the physical ones; Our numerical results are presented in 
Tabs.(\ref{tab1},\ref{tab1b}). Under this hypotheese, the mixing parameter is zero, see 
Eq.(\ref{bgmcc}), then the bosons $W$ and $V$ are the physical ones. In this 
scenario we can easily conclude the following hierarque in the masses are
\begin{equation}
M_{Z^{\prime}}>M_{U}>M_{V},
\label{generalmassrel}
\end{equation}
this result is in agreement with the previous analysis presented in \cite{Rodriguez:2010tn}.
\item[b-)] Consider all new vev are non zero, where $W_{1}$ and $W_{2}$ are the physical gauge 
bosons. Our results can be found in Tabs.(\ref{tab2},\ref{tab2b}). In this case we get the 
following results
\begin{itemize}
\item $M_{W_{1}}\cong M_{W}$ and $M_{W_{2}}\cong M_{V}$;
\item $M_{U}=468.476$ GeV and $M_{Z^{\prime}}=2023.910$ GeV.
\end{itemize}
The Eq.(\ref{generalmassrel}) is still hold in this case.
\end{itemize}

\begin{table}
\begin{tabular}{|c|c|c|c|c|}
\hline
$v_{\sigma^{\prime}_{2}}$ GeV & $M_{U}$ GeV & $M_{V}$ GeV & $M_{Z^{\prime}}$ GeV 
& $\delta_{ZZ^{\prime}}$ GeV$^{2}$  \\ 
\hline
12.0 & 468.506 & 461.584 & 2023.910 & -21956.1 \\
12.5 & 468.512 & 461.586 & 2023.910 & -21956.0 \\
13.0 & 468.518 & 461.587 & 2023.910 & -21955.8 \\
13.5 & 468.524 & 461.589 & 2023.910 & -21955.6 \\
14.0 & 468.530 & 461.590 & 2023.910 & -21955.5 \\
14.5 & 468.536 & 461.592 & 2023.910 & -21955.3 \\
15.0 & 468.543 & 461.593 & 2023.910 & -21955.1 \\
15.5 & 468.550 & 461.595 & 2023.910 & -21954.9 \\
16.0 & 468.557 & 461.597 & 2023.910 & -21954.7 \\
\hline
\end{tabular}
\caption{Masses of new gauge bosons for some values of 
$v_{\sigma^{\prime}_{2}}$ but $v_{\sigma_{1}}=v_{\sigma^{\prime}_{1}}=0$ GeV; 
$v_{\chi}=v_{\chi^{\prime}}=1000$ GeV and the values of $M_{V}$, see Eq.(\ref{explimV}), and 
the values 
of $M_{Z^{\prime}}$, Eq.(\ref{explimZP}) and they satisfy the bounds presented by 
Eqs.(\ref{explimV},\ref{explimZP}).}
\label{tab1}
\end{table}

\begin{table}
\begin{tabular}{|c|c|c|c|}
\hline
$v_{\chi^{\prime}}$ GeV & $M_{U}$ GeV & $M_{V}$ GeV & $M_{Z^{\prime}}$ GeV \\ 
\hline
1000 & 468.498  & 461.582  & 2023.91  \\
1500 & 593.809  & 588.368  & 2579.43  \\
2000 & 734.150  & 729.757  & 3199.01  \\
2500 & 882.379  & 878.726  & 3851.85  \\
3000 & 1035.110 & 1032.000  & 4523.57  \\
3500 & 1190.620 &  1187.910 & 5206.88  \\
4000 & 1347.930 & 1345.550  & 5897.75  \\
4500 & 1506.500 & 1504.360 & 6593.79  \\
5000 & 1665.950 & 1664.020   & 7293.54  \\
\hline
\end{tabular}
\caption{Masses for all new gauge bosons of this model for some values of $v_{\chi^{\prime}}$ 
with $v_{\sigma^{\prime}_{2}}=11.1942$ GeV; $v_{\chi}=1000$ GeV;  $v_{\sigma_{1}}=v_{\sigma^{\prime}_{1}}=0$ GeV and they are in agreement with the bounds at Eqs.(\ref{explimV},\ref{explimZP}). The values for $\delta_{ZZ^{\prime}}=-21956.4$ GeV$^{2}$.}
\label{tab1b}
\end{table}

\begin{table}
\begin{tabular}{|c|c|c|c|}
\hline
$v_{\sigma_{1}}$ GeV & $M_{W}$ GeV & $M_{V}$ GeV & $\delta_{WV}$ GeV$^{2}$ \\ 
\hline
0.0 & 80.4332 & 461.582 & 12.9285 \\
0.5 & 80.4335 & 461.582 & 14.4345  \\
1.0 & 80.4345 & 461.582 & 15.9406 \\
1.5 & 80.4361 & 461.582 & 17.4467  \\
2.0 & 80.4385 & 461.583 & 18.9528 \\
2.5 & 80.4414 & 461.583 & 20.4588 \\
3.0 & 80.4451 & 461.584 & 21.9649 \\
3.5 & 80.4494 & 461.585 & 23.4710 \\
4.0 & 80.4544 & 461.586 & 24.9771 \\
4.5 & 80.4600 & 461.587 & 26.4831 \\
5.0 & 80.4663 & 461.588 & 27.9892 \\
5.5 & 80.4732 & 461.589 & 29.4953 \\
\hline
\end{tabular}
\caption{Masses of single charged gauge bosons and $\delta_{WV}$ for some 
values of $v_{\sigma_{1}}$ for $v_{\chi}=v_{\chi^{\prime}}=1000$ GeV;  $v_{\sigma^{\prime}_{1}}=4.89453$ GeV; 
$v_{\sigma^{\prime}_{2}}=8.7691841$ GeV.}
\label{tab2}
\end{table}

\begin{table}
\begin{tabular}{|c|c|c|c|c|}
\hline
$v_{\chi^{\prime}}$ GeV & $M_{W}$ GeV & $M_{V}$ GeV & $M_{W_{1}}$ GeV & $M_{W_{2}}$ GeV \\ 
\hline
1000 & 80.4335 & 461.582 & 80.4335  & 461.582 \\
1500 & 80.4335 & 588.368 & 80.4335  & 588.368 \\
2000 & 80.4335 & 729.757 & 80.4335  & 729.757 \\
2500 & 80.4335 & 878.726 & 80.4335  & 878.726 \\
3000 & 80.4335 & 1032.000 & 80.4335  & 1032.000 \\
3500 & 80.4335 & 1187.91 & 80.4335  & 1187.91 \\
4000 & 80.4335 & 1345.55 & 80.4335  & 1345.55 \\
4500 & 80.4335 & 1504.36 & 80.4335  & 1504.36 \\
5000 & 80.4335 & 1664.02 & 80.4335  & 1664.02 \\
\hline
\end{tabular}
\caption{Masses of $M_{W},M_{V}, M_{W_{1}}$ and $M_{W_{2}}$, see Eq.(\ref{thetac}), for some 
values of $v_{\chi^{\prime}}$ for $v_{\chi}=1000$ GeV; $v_{\sigma_{1}}0.5$ GeV; $v_{\sigma^{\prime}_{1}}=4.89453$ GeV; 
$v_{\sigma^{\prime}_{2}}=8.7691841$ GeV.}
\label{tab2b}
\end{table}

\section{Conclusions}
\label{sec:con}

We show that in the context of Minimal Supersymmetric 
$SU(3)_{C}\otimes SU(3)_{L}\otimes U(1)_{N}$ Model, MSUSY331, 
we can explain both the new measurement on the $W$-boson 
mass as well as the $\rho$-parameter if 
Eqs.(\ref{wmassrest},\ref{rhorest}) are satisfied 
simultaneously as we have showed in Eqs.(\ref{mzhip1},\ref{mzhip3}) 
and also in our Fig.(\ref{fig6}). We also calculated the masses of all the gauge bosons of 
this model and $M_{Z^{\prime}}>M_{U}>M_{V}$ and their values are in agreement with the 
actual experimental data given at Eqs.(\ref{explimV},\ref{explimZP}).

\begin{center}
{\bf Acknowledgments} 
\end{center}
We would like thanks V. Pleitez for useful discussions above 331 models and 
about this intersting research topic. We also to thanks IFT for the nice hospitality during 
my several visit to perform my studies about the severals 331 Models and also for done 
this article.

\appendix

\section{Mixing between $W$ and $V$.}
\label{sec:mixingchar}

We notice, from Eq.(\ref{bgmcc}), 
if $v_{\sigma_{1}}\neq 0$ the 
gauge bosons $W$ and $V$ are no longer the physical ones, in 
agreement with the results presented at \cite{Liu:1993gy,Montero:1999mc,VanLoi:2022eir}. They 
can mixing and the physical bosons are $W^{\pm}_{1,2}$ and their masses are
\begin{eqnarray}
M^{2}_{W_{1}}&=& \frac{1}{2} \left( 
M^{2}_{W}+M^{2}_{V}- \sqrt{(M^{2}_{W}-M^{2}_{V})^{2}+4 \delta^{2}_{WV}} 
\right), \nonumber \\
M^{2}_{W_{2}}&=& \frac{1}{2} \left( 
M^{2}_{W}+M^{2}_{V}+ \sqrt{(M^{2}_{W}-M^{2}_{V})^{2}+4 \delta^{2}_{WV}} 
\right).
\label{eigenvaluescc}
\end{eqnarray}
The physical eigenstates are defined as
\begin{eqnarray} 
\left( \begin{array}{c}
W^{\pm}_{1m}\\
W^{\pm}_{2m} \end{array} \right)= \left( \begin{array}{cc}
\cos \theta^{\pm}&- \sin \theta^{\pm}\\
\sin \theta^{\pm}& \cos \theta^{\pm} 
\end{array} 
\right) \left( 
\begin{array}{c}
W^{\pm}_{m}\\
V^{\pm}_{m} \end{array}\right),
\label{eigenvectorscc}
\end{eqnarray}
the mixing angle is given by
\begin{equation}
\tan^{2} \theta^{\pm}= 
\frac{M^{2}_{ W_{1}}-M^{2}_{W}}{M^{2}_{ W_{2}}-M^{2}_{V}}.
\label{thetac}
\end{equation}

\section{Mixing between $Z$ and $Z^{\prime}$.}
\label{sec:mixingneu}

The neutral gauge bosons $Z$ and $Z^{\prime}$ can mix and it 
is given by
\begin{eqnarray} 
{\cal L}^{\mbox{neutra}}&=& \left( \begin{array}{cc}
Z_{m}& Z^{\prime}_{m} \end{array} \right) \left( \begin{array}{cc}
M^{2}_{Z}& \delta_{ZZ^{\prime}} \\
\delta_{ZZ^{\prime}}& M^{2}_{Z^{\prime}} \end{array} \right) \left( \begin{array}{c}
Z_{m}\\
Z^{\prime}_{m} \end{array} \right),
\label{bgmcc}
\end{eqnarray}
where 
\begin{eqnarray}
M^{2}_{Z}&=& \frac{g^{2}v^{2}_{MP}}{4 \cos^{2} \theta_{W}}, \nonumber \\
M^{2}_{Z^{\prime}}&=& \frac{g^{2}}{4}\left[ \frac{4v^{2}_{MP}}{3s}+ \frac{2t^{2}U^{2}}{s}+ \frac{4s}{3} 
\left( w^{2}+ w^{\prime 2} \right) \right], \nonumber \\
\delta_{ZZ^{\prime}}&=& \frac{g^{2}}{4 \sqrt{3h_{w}}s}\left[ (V^{2}-U^{2})-6t^{2}U^{2} \right],
\label{zmass}
\end{eqnarray}
compare those values with our result presented on Eqs.(\ref{zmassmcr}). We have defined
\begin{eqnarray}
V^{2}&=&v^{2}+2y^{2}+z^{2}+v^{\prime 2}+2y^{\prime 2}+z^{\prime 2}, \nonumber \\
U^{2}&=&u^{2}+ u^{\prime 2}, \nonumber \\
t^{2}&=& \frac{\sin^{2}\theta_{W}}{1-4 \sin^{2}\theta_{W}}, \nonumber \\
s&=&1+3t^{2}, \nonumber \\
v^{2}_{MP}&=&V^{2}+U^{2},
\end{eqnarray}
to understand how to get the third expression above see Eqs.(\ref{tinthetaw},\ref{tinthetaw2}). 
We can show the following relations
\begin{eqnarray}
\cos \theta_{W}&=& \frac{\sqrt{s}}{\sqrt{s+t^{2}}}, \nonumber \\
\sin \theta_{W}&=& \frac{t}{\sqrt{s+t^{2}}}, \nonumber \\
\sqrt{s+t^{2}}&=&\left( \sqrt{h_{W}} \right)^{-1}, \nonumber \\
h_{W}&=&1-4 \sin^{2}\theta_{W}.
\label{def:hw}
\end{eqnarray}

The physical mass of physical gauge bosons are
\begin{eqnarray} 
\left( \begin{array}{c}
Z^{0}_{1m}\\
Z^{0}_{2m} \end{array} \right)= \left( \begin{array}{cc}
\cos \theta^{0}&- \sin \theta^{0}\\
\sin \theta^{0}& \cos \theta^{0} 
\end{array} 
\right) \left( 
\begin{array}{c}
Z^{0}_{m}\\
(Z^{\prime})^{0}_{m} \end{array}\right),
\label{eigenvectorscc}
\end{eqnarray}
where the mixing angle in this sector is
\begin{equation}
\tan^{2} \theta^{0}= 
\frac{M^{2}_{ Z_{1}}-M^{2}_{Z}}{M^{2}_{ Z_{2}}-M^{2}_{Z^{\prime}}},
\label{thetac}
\end{equation}
where
\begin{eqnarray}
M^{2}_{Z_{1}}&=& \frac{1}{2} \left( 
M^{2}_{Z}+M^{2}_{Z^{\prime}}+ \sqrt{(M^{2}_{Z}-M^{2}_{Z^{\prime}})^{2}+4 
\delta^{2}_{ZZ^{\prime}}} 
\right), \nonumber \\
M^{2}_{Z_{2}}&=& \frac{1}{2} \left( 
M^{2}_{Z}+M^{2}_{Z^{\prime}}- \sqrt{(M^{2}_{Z}-M^{2}_{Z^{\prime}})^{2}+4 
\delta^{2}_{ZZ^{\prime}}} 
\right).
\label{eigenvaluesnc}
\end{eqnarray}

\section{Preliminar Numerical Analysis for $Z$ and $Z^{\prime}$}
\label{zezp}

For our numerical analyses, at tree-level, we will apply the constraint\footnote{As we have choose at \cite{331susy1}.} 
\begin{equation}
V^{2}_{\eta}+V^{2}_{\rho}+2V^{2}_{2}=(246\;{\rm GeV})^{2},
\end{equation} 
coming from $M_{W}$, where,
we have defined 
\begin{eqnarray}
V^{2}_{\eta}&=&v^{2}_{\eta}+v^{2}_{ \eta^{\prime}}, \,\ 
V^{2}_{\rho}=v^{2}_{\rho}+v^{2}_{ \rho^{\prime}}, \nonumber \\ 
V^{2}_{2}&=&v^{2}_{\sigma_{2}}+v^{2}_{\sigma^{\prime}_{2}}.
\end{eqnarray}
Assuming that 
\begin{equation}
v_{\eta}=20, \,\ v_{\sigma_{2}}=10, \,\ 
v_{ \eta^{\prime}}=v_{ \rho^{\prime}}=1 \,\ {\mbox GeV},
\label{vevescolha}
\end{equation}
The value of $v_{\rho}$ is fixed by Eq.(\ref{smvalue}), we get
\begin{equation}
v_{\rho}=245.20 \,\ {\mbox GeV}.
\end{equation}

As an example as the mixing of $Z$ and $Z^{\prime}$ we use the vev defined in 
Eq.(\ref{vevescolha}) together the following vev
\begin{eqnarray}
v_{\sigma_{1}}&=&v_{\sigma^{\prime}_{1}}=
v_{\sigma^{\prime}_{2}}=0 \,\ {\mbox GeV}, \nonumber \\
v_{\chi}&=&v_{\chi^{\prime}}=1000 \,\ {\mbox GeV},
\end{eqnarray}
in Eq.(\ref{zmass}) to get the following results\footnote{The 
parameter $\delta_{ZZ^{\prime}}$ has no dependence in 
$v_{\chi}$ and $v_{\chi^{\prime}}$, see Eq.(\ref{zmass}), do it 
we will shown only this specific case, see also our Tabs.(\ref{tab1},\ref{tab1b}).}
\begin{eqnarray}
M_{Z}&=&91.1875 \,\ {\mbox GeV}, \,\
M_{Z^{\prime}}=2023.91 \,\ {\mbox GeV}, \nonumber \\
\delta_{ZZ^{\prime}}&=&-21958 \,\ {\mbox GeV}^{2},
\label{ZSM}
\end{eqnarray}
the value for $Z^{\prime}$ mass is in agreement with Eq.(\ref{explimZP}). Repare $\delta_{ZZ^{\prime}} \neq 0$, therefore $Z$ and 
$Z^{\prime}$ are not physical gauge boson. The masses of 
physical gauge bosons $Z_{1,2}$, using 
Eq.(\ref{eigenvaluesnc}), are\footnote{Compare 
$\delta_{ZZ^{\prime}}$ with $M^{2}_{Z}$ and 
$M^{2}_{Z^{\prime}}$ and it is smaller therefore the mixing 
parameter is almost negligible.}
\begin{eqnarray}
M_{Z_{1}}&=&91.1875 \,\ {\mbox GeV}, \,\
M_{Z_{2}}=2023.91 \,\ {\mbox GeV}.
\end{eqnarray}
Therefore the bosons $Z$ and $Z^{\prime}$, as a first approximation, can be considered 
as being the physical states, and their masses given at Eq.(\ref{zmass}).



\begin{thebibliography}{99}
\bibitem{cdf}T. Aaltonen {\it et al} (CDF Collaboration), 
{\it High-precision measurement of the $W$ boson mass with the CDF II 
detector}, {\sl Science} {\bf 376}, 170, (2022).
\bibitem{sg}S. J. L. Rosner,
{\it Resource letter: The Standard model and beyond},
{\sl Am. J. Phys.} {\bf 71}, 302, (2003).
\bibitem{Kronfeld:2010bx} A. S. Kronfeld and C. Quigg,
{\it Resource Letter: Quantum Chromodynamics},
{\sl Am. J. Phys.}{\bf78}, 1081, (2010).


\bibitem{average}J. de Blas, M. Pierini, L. Reina and L. Silvestrini, 
{\it Impact of the recent measurements of the top-quark and $W$-boson masses 
on electroweak precisions fits}, [arXiv:2204.04204].

\bibitem{pp} F. Pisano and V. Pleitez, 
{\it An $SU(3)\otimes U(1)$ model for electroweak interactions},
{\sl Phys. Rev.}{\bf D46}, 410, (1992).

\bibitem{pf} R. Foot, O. F. Hernandez, F. Pisano and V. Pleitez, 
{\it Lepton masses in an $SU(3)_{L} \otimes U(1)_{N}$ gauge model},
{\sl Phys. Rev.} {\bf D 47}, 4158, (1993).

\bibitem{VanLoi:2022eir} D. Van Loi and P. Van Dong,
{\it Novel effects of the W-boson mass shift in the 3-3-1 model},
{\sl Eur. Phys. J.}{\bf C83}, 56, (2023); [arXiv:2206.10100 [hep-ph]].
\bibitem{Pisano:1997hk} F. Pisano and S. S. Sharma,
{\it Majoron emitting neutrinoless double beta decay in the electroweak chiral gauge extensions},
{\sl Phys. Rev.}{\bf D57}, 5670, (1998).



\bibitem{ema1} T. V. Duong and E. Ma, 
{\it Supersymmetric $SU(3) \otimes U(1)$ Gauge Model: 
Higgs Structure at the Electroweak Energy Scale}, 
{\sl Phys. Lett.}{\bf B316}, 307 (1993).
\bibitem{pal2} H. N. Long and P. B. Pal, {\it Nucleon instability in a supersymmetric 
$SU(3)_{C}\otimes SU(3)_{L}\otimes U(1)$ model}, 
{\sl Mod. Phys. Lett.}{\bf A13}, 2355, (1998).


\bibitem{331susy1}J. C. Montero, V. Pleitez and M. C. Rodriguez, 
{\it A Supersymmetric 3-3-1 model},
{\sl Phys. Rev.} {\bf D65}, 035006, (2002).

\bibitem{mcr} M. Capdequi-Peyran\`ere and M.C. Rodriguez, 
{\it Charginos and neutralinos production at 3-3-1 supersymmetric model in e- e- scattering},
{\sl Phys. Rev.} {\bf D 65}, 035001 (2002).
\bibitem{Montero:2001ch}
J. C. Montero, V. Pleitez and M. C. Rodriguez,
{\it Lepton masses in a supersymmetric 3-3-1 model},
{\sl Phys. Rev.}{\bf D65}, 095008, (2002).



\bibitem{Liu:1993gy}
J. T. Liu and D. Ng,
{\it Lepton flavor changing processes and CP violation in the 331 model},
{\sl Phys. Rev.}{\bf D50}, 548, (1994), [arXiv:hep-ph/9401228 [hep-ph]].
\bibitem{Montero:1999mc}
J. C. Montero, C. A. de Sousa Pires and V. Pleitez,
{\it Spontaneous breaking of a global symmetry in a 331 model},
{\sl Phys. Rev.}{\bf D60}, 115003, (1999),
[arXiv:hep-ph/9903251 [hep-ph]].
\bibitem{DeConto:2015eia} G. De Conto, A. C. B. Machado and V. Pleitez,
{\it Minimal 3-3-1 model with a spectator sextet},
{\sl Phys. Rev.}{\bf D92}, 075031, (2015), [arXiv:1505.01343 [hep-ph]].
\bibitem{Gelmini:1980re} G. B. Gelmini and M. Roncadelli,
{\it Left-Handed Neutrino Mass Scale and Spontaneously Broken Lepton Number},
{\sl Phys. Lett.}{\bf B99}, 411 (1981).
\bibitem{Rodriguez:2021dnb}M. C. Rodriguez,
{\it The Scalar Potential of Supersymmetric $SU(3)_{C}\otimes SU(3)_{L}\otimes U(1)_{N}$ Model}, [arXiv:2111.15393 [hep-ph]].
\bibitem{Rodriguez:2005jt}M. C. Rodriguez,
{\it Scalar sector in the minimal supersymmetric 3-3-1 model},
{\sl Int. J. Mod. Phys.}{\bf A21}, 4303, (2006).
\bibitem{Rodriguez:2010tn}M. C. Rodriguez,
{\it Mass Spectrum in the Minimal Supersymmetric 3-3-1 model},
{\sl J. Mod. Phys.}{\bf 2}, 1193, (2011).
\bibitem{osland} P. Osland, A. A. Pankov and I. A. Serenkova, 
{\it Bounds on the mass and mixing of $Z^{\prime}$ and 
$W^{\prime}$ bosons decaying into different pairings of 
$W$,$Z$, or Higgs bosons using CMS data at the LHC},
[arXiv:2206.01438] [hep-ph].
\bibitem{Dias:2004dc}
A. G. Dias, R. Martinez and V. Pleitez,
{\it Concerning the Landau pole in 3-3-1 models},
{\sl Eur. Phys. J.}{\bf C39}, 101, (2005); [arXiv:hep-ph/0407141 [hep-ph]].
\bibitem{Montero:1999su}
J. C. Montero, C. A. de Sousa Pires and V. Pleitez,
{\it Comment on `Majoron emitting neutrinoless double beta decay in the electroweak chiral gauge extensions'},
{\sl Phys. Rev.}{\bf 60}, 098701, (1999).
\bibitem{Sharma:1999ct} S. S. Sharma and F. Pisano,
{\it Reply to `comment on: 'Majoron emitting neutrinoless double beta decay in the electroweak chiral gauge extensions'},
{\sl Phys. Rev.}{\bf 60}, 098702, (1999).


\bibitem{Blank:1997qa}T. Blank and W. Hollik,
{\it Precision observables in SU(2) x U(1) models with an additional Higgs triplet},
{\sl Nucl. Phys.}{\bf B514}, 113-134, (1998);
[arXiv:hep-ph/9703392 [hep-ph]].
\bibitem{Chen:2006pb}M. C. Chen, S. Dawson and T. Krupovnickas,
{\it Higgs triplets and limits from precision measurements},
{\sl Phys. Rev.}{\bf D74}, 035001, (2006);
[arXiv:hep-ph/0604102 [hep-ph]].
\bibitem{evans}J. L. Evans, T. T. Yanagida and N. Yokozaki,
{\it W boson mass anomaly and grand unification},
[arXiv:2205.03887] [hep-ph].

\bibitem{Kribs:2007ac} G. D. Kribs, E. Poppitz, and N. Weiner, 
{\it Flavor in supersymmetry with an extended R-symmetry},
{\sl Phys. Rev.} {\bf D78}, 055010, (2008).
\bibitem{Diessner:2014ksa}P. Diessner, J. Kalinowski, W. Kotlarski, and D. St\"ockinger, 
{\it Higgs boson mass and electroweak observables in the MRSSM},
{\sl JHEP} {\bf 12}, 124, (2014).
\bibitem{Diessner:2019ebm} P. Diessner and G. Weiglein, 
{\it Precise prediction for the W boson mass in the MRSSM}, 
{\sl JHEP} {\bf 07}, 011, (2019).
\bibitem{athron}P. Athron, M. Bach, D. H. J. Jacon, W. Kotlarski, D. 
St\"ockinger and A. Voigt, {\it Precise calculation of the $W$ boson pole 
mass beyond the Standard Model with FlexibleSUSY}, [arXiv:2204.05285] [hep-ph].















\end{thebibliography}
\end{document}